\documentclass[a4paper,11pt]{article}
\usepackage{pos}
    \usepackage{lineno}
    \usepackage{float}

\title{Measurement of Two-Point Energy Correlators Within Jets in $pp$ Collisions at $\sqrt{s}$ = 200 GeV at STAR}
\author*{Andrew Tamis, for the STAR Collaboration}

\affiliation{Wright Lab, Yale University,\\
  New Haven, CT, United States}

\emailAdd{andrew.tamis@yale.edu}

\abstract{Jet substructure is a powerful tool to probe the time evolution of a parton shower. However, many of the analysis methods used to extract splitting formation times from jet substructure, such as Soft Drop grooming and the Lund plane, focus on the hardest radiation of the jet. A complementary observable with growing theoretical and experimental interest, the 2-point Energy Correlator (EEC), re-contextualizes jet substructure study by using the distribution of angular distance of all combinations of two final state particles within a jet. This distribution is weighted by the product of the fractions of jet energy that each of the constituents carry, and thus is infrared-and-collinear safe. The EEC can reveal the separation between two distinct regimes: effects originating from free hadrons at small opening angles and from perturbative fragmentation of quarks and gluons at large opening angles.

In these proceedings, the first measurement of the EEC at RHIC is presented, using the data taken at $\sqrt{s}$ = 200 GeV $pp$ collisions by STAR. The EEC will be shown for several full jet transverse momentum selections and compared to predictions from the PYTHIA-8 Detroit tune. This study is useful as a baseline for comparisons to future studies in heavy-ion collisions, which will provide information about how the quark-gluon plasma interacts with the jet across different angular scales.}

\FullConference{HardProbes2023\\
 26-31 March 2023 \\
 Aschaffenburg, Germany\\}

\begin{document}
\maketitle
\nolinenumbers

\section{Introduction}
Hard-scattered partons in high energy collisions undergo angular-ordered fragmentation and ultimately hadronize into final-state particles that are then measured by detectors.  These final-state particles are then clustered, using jet clustering algorithms, in order to define the experimental signal of a jet - a proxy for the initial hard-scattered parton.  The clustering of the constituents within a jet, its substructure, allows for the study of information encoded during the fragmentation and hadronization processes.

Many jet substructure observables, such as those utilizing SoftDrop grooming \cite{softDrop}, isolate the time information of a jet by focusing on only the hardest constituents and splittings: which corresponds to perturbative quantum chromodynamics (QCD) effects.  However, of particular interest for studying the limits of perturbative QCD is the region where non-perturbative effects begin to dominate as partons are confined into hadrons.  N-point energy correlation functions have long been proposed in theory and previously studied in electron-positron collisions \cite{TOPAZ:1989yod}, but there has been recent interest in applying them to jets produced in $pp$ and heavy-ion collision systems \cite{Chen:2020vvp}, taking advantage of advancements in understanding of jet substructure.  The 2-point energy correlator (EEC) \cite{Komiske:2022enw} aims to re-contextualize jet substructure study by using all charged constituents within a jet, looking at the distribution of energy between all combinations of two of them plotted differentially in their angular separation in azimuthal-angle ($\phi$) and pseudorapidity ($\eta$) space: $\Delta R = \sqrt{\Delta\eta^2+\Delta\phi^2}$.  This observable separates the angular distribution of the jet constituents into three regimes: scaling corresponding to the diffusion of non-perturbative hadrons at low angles and behavior corresponding to the perturbative fragmentation of the parton shower at large opening angles, separated by a transition region between them.  By relating this opening angle to formation time, $t_f \approx {1}/{\Delta R ^2}$ \cite{Apolinario:2022kgo}, it is possible to interpret these regions in terms of time: with fragmentation occurring earlier in time and hadronization later.  The first experimental measurement of EECs at STAR is presented in these proceedings for several selections on jet transverse momentum ($p_{\rm T}$).

\section{Experimental Details}

This analysis was done using the data from $pp$ collisions at $\sqrt{s}$ = 200 GeV recorded by the STAR experiment \cite{STAR} in 2012.  Charged tracks are reconstructed using the Time Projection Chamber (TPC).  Neutral energy deposits are determined using the Barrel Electro-Magnetic Calorimeter (BEMC).  Events are selected using a jet trigger which requires an energy deposit of at least 7.3 GeV in a BEMC patch 1x1 in $\eta-\phi$ space.  Jets are reconstructed using the anti-$k_T$ jet-finding algorithm with resolution parameters R = 0.4 and 0.6 \cite{Cacciari:2011ma}. Both charged tracks found via the TPC together with neutral energy deposited in the BEMC are used in the jet finding and in the determination of the jet momentum.

The EEC is a weighted distribution of the products of the jet energy fractions carried by all possible two-constituent combinations within a jet.  For the purposes of this analysis, this was done using only charged tracks reconstructed in the STAR TPC, in order to take advantage of its excellent tracking resolution for angular distances.  

The experimental definition of the two-point energy correlator used in this analysis is given by:

\begin{equation}
\text{Normalized EEC} = \frac{1}{\Sigma_{jets}\Sigma_{i \neq j}\frac{E_i E_j}{p_{\rm T,jet}^2}} \frac{d(\Sigma_{\rm jets}\Sigma_{i \neq j}\frac{E_i E_j}{p_{\rm T,jet}^2})}{d \Delta R}
\label{EEC_equation}
\end{equation}

where $E_i$ represents the energy of the $\rm i^{th}$ constituent within a jet, i and j are two distinct constituents that make up a particle pair and $\Delta R$ is the angular distance between the two particles in the pair.  This is effectively a cross section of the number of charged particle pairs, differential in their angular distance.  However, each pair is scaled by an additional energy weight, ${E_i E_j}/{p_{\rm T,jet}^2}$, in order to suppress soft radiation without removing it completely, making this observable infrared and collinear safe.  Importantly, this allows for direct comparison to theory calculations in the perturbative regime.  The energy of each constituent is determined via its four-momentum, assuming a pion mass.  The integral of this distribution is then normalized to unity within each $p_{\rm T,jet}$ bin, in order to more easily compare any potential shape differences.

\begin{figure}[H]
\centering
    \includegraphics[width=0.46\linewidth]{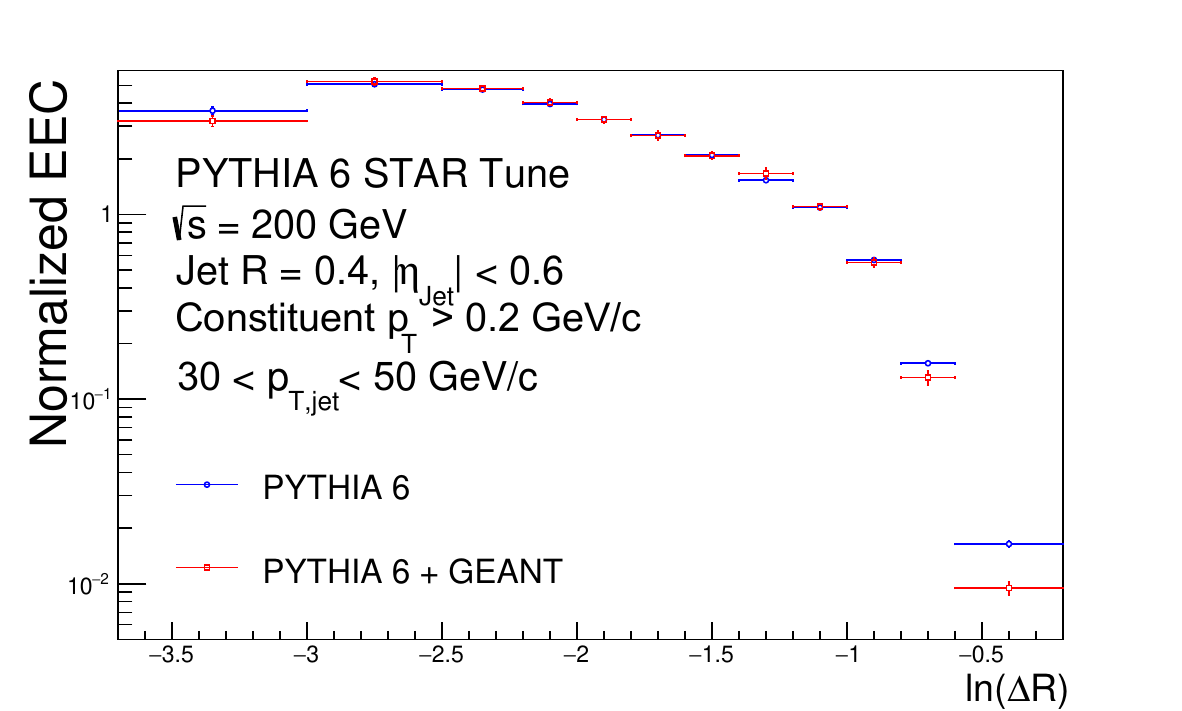}
\includegraphics[width=0.46\linewidth]{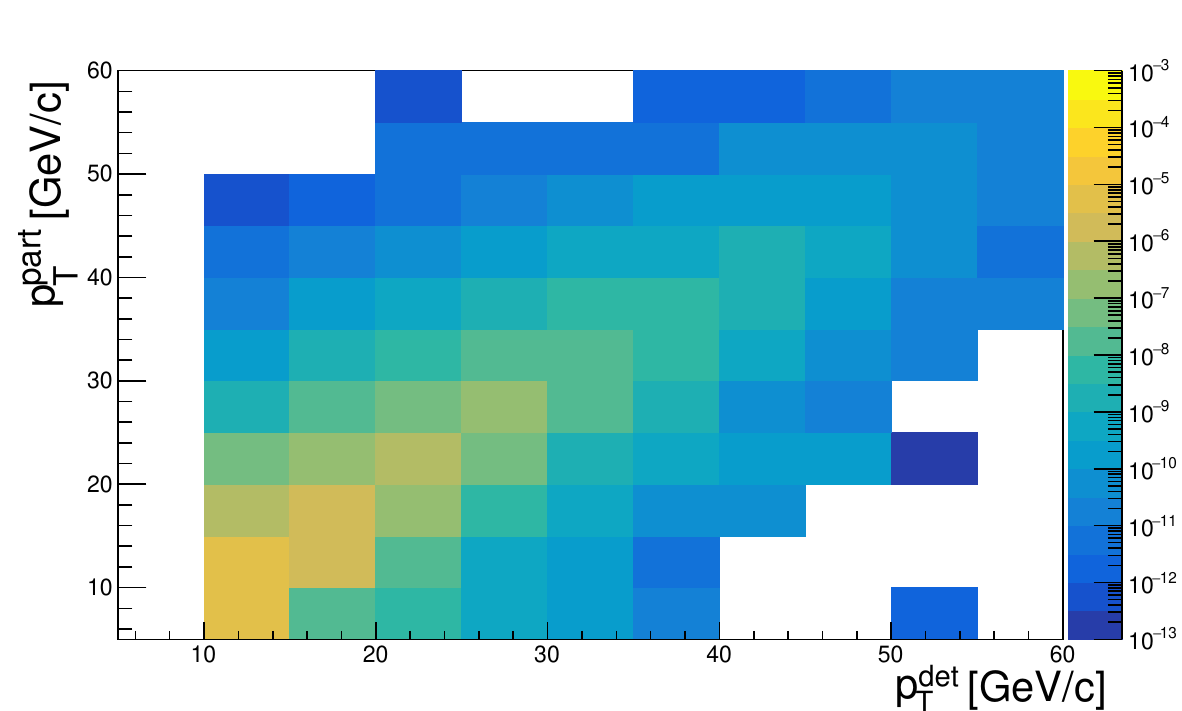}

\caption{A comparison between the PYTHIA-6 (truth level) and the GEANT embedded sample (particle level) EEC (left). Response Matrix showing mapping from detector-level transverse jet momentum ($p_{\rm T}^{\rm{\rm{det}}}$) to particle-level ($p_{\rm T}^{\rm{\rm{part}}}$) calculated using particle pairs found in matched jets of these two simulation samples (right).}
\label{Response}
\end{figure}

As this observable is sensitive to selections on $p_{\rm T,jet}$, the distribution of which is sensitive to detector effects, correction must be performed in order to reconstruct the true $p_{\rm T,jet}$.  This correction is performed by comparing a simulation at the particle (truth) level and detector level.  This was accomplished by using a particle-level sample generated via the Monte-Carlo event generator PYTHIA-6 \cite{Sjostrand:2006za} and passing it through GEANT3 \cite{Brun:1994aa}, a full simulation of the STAR experiment to create a simulated detector-level sample.  A comparison of the EEC between these two samples for a selection of 30 < $p_{\rm{T, jet}}$ < 50 GeV/c can be seen in Fig. \ref{Response} on the left, showing that the overall needed correction is small, and that the largest discrepancies occur at very low and very high $\Delta$R.  Jets are matched between the two samples geometrically if their jet axes are separated by less than the jet radius.  Likewise, charged tracks within matched jets are matched with a matching radius of 0.02 radians.  Using these two samples, a response matrix, shown on the right in Fig. \ref{Response}, is constructed: mapping the detector jet transverse momentum $p_{\rm T}^{\rm{det}}$ to the truth transverse momentum $p_{\rm T}^{\rm part}$ for each pair.  Once this matrix is constructed, a selection in $p_{\rm T}^{\rm{part}}$ is taken to produce a fractional weight corresponding to each $p_{\rm T}^{\rm{det}}$ bin.  The normalized EEC distributions for each measured bin taken from the STAR data are then added in these fractions in order to reproduce the truth level EEC distribution for a given $p_{\rm T}^{\rm{part}}$ bin.  Additionally, it is possible for correlations from the particle-level sample to not be found in the detector-level sample for several reasons.  For example, constituents of selected particle-level jets may be lost due to tracking efficiency. These misses are added back into the distribution from the PYTHIA-6 STAR Tune \cite{STAR_Tune}.

Several systematic uncertainties were determined by varying the properties of the GEANT3 simulation, including the hadronic correction used in data analysis, the tower gain, and the tracking efficiency of the TPC.  Similar uncertainties were used in previous studies done by STAR \cite{STAR:2020ejj,STAR:2021lvw}.  In addition to the effect of varying the properties of the detector simulation, the additional uncertainty of the variables that were not corrected for -- i.e. the $\Delta$R and energy weight ($E_1 E_2/ p_{\rm T, jet}$), were accounted for as the maximum difference in the distribution for matched jets selected on $p_{\rm T}^{\rm det}$ between the truth and detector-level samples, which was seen to be the dominant systematic on the order of 5$\%$.

\section{Results}

\begin{figure}[ht]
\centering
\includegraphics[width=0.49\linewidth]{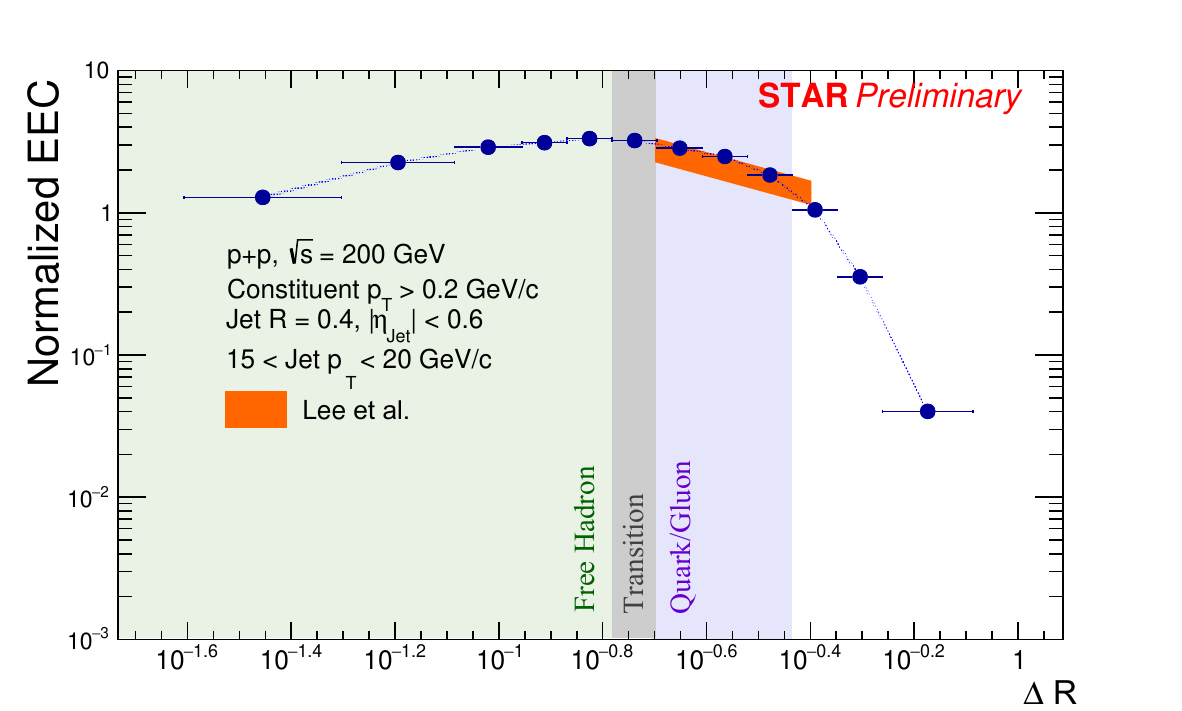}
\includegraphics[width=0.49\linewidth]{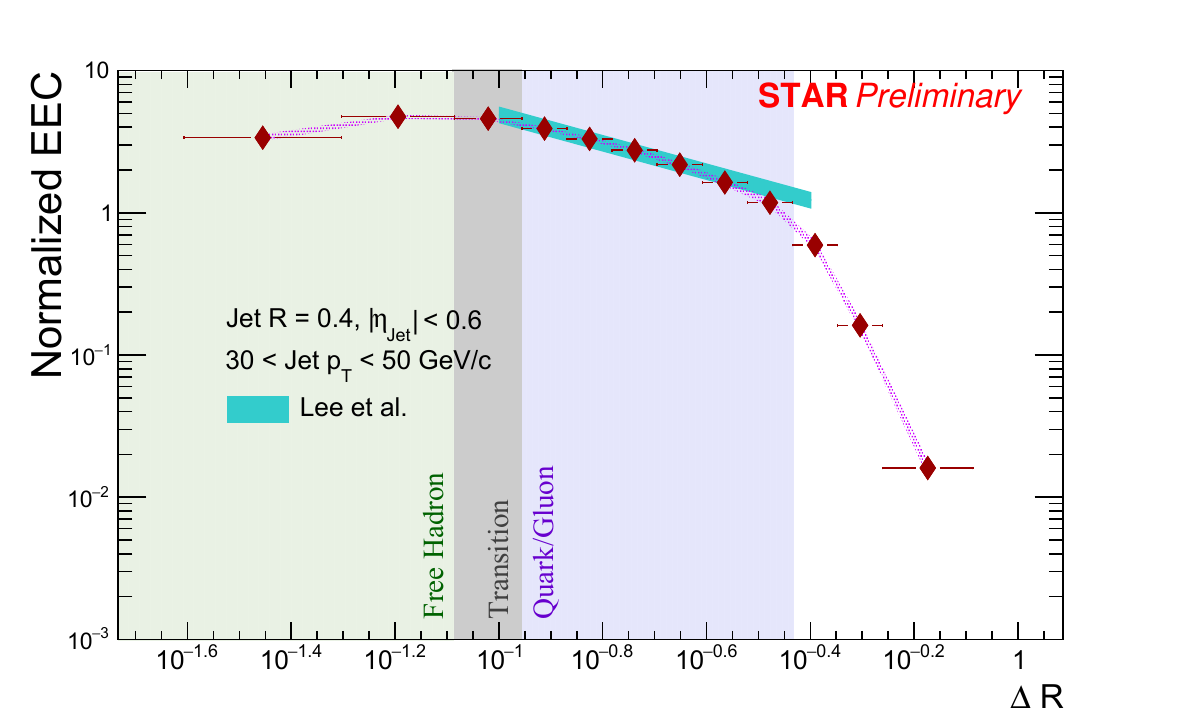}
\includegraphics[width=0.49\linewidth]{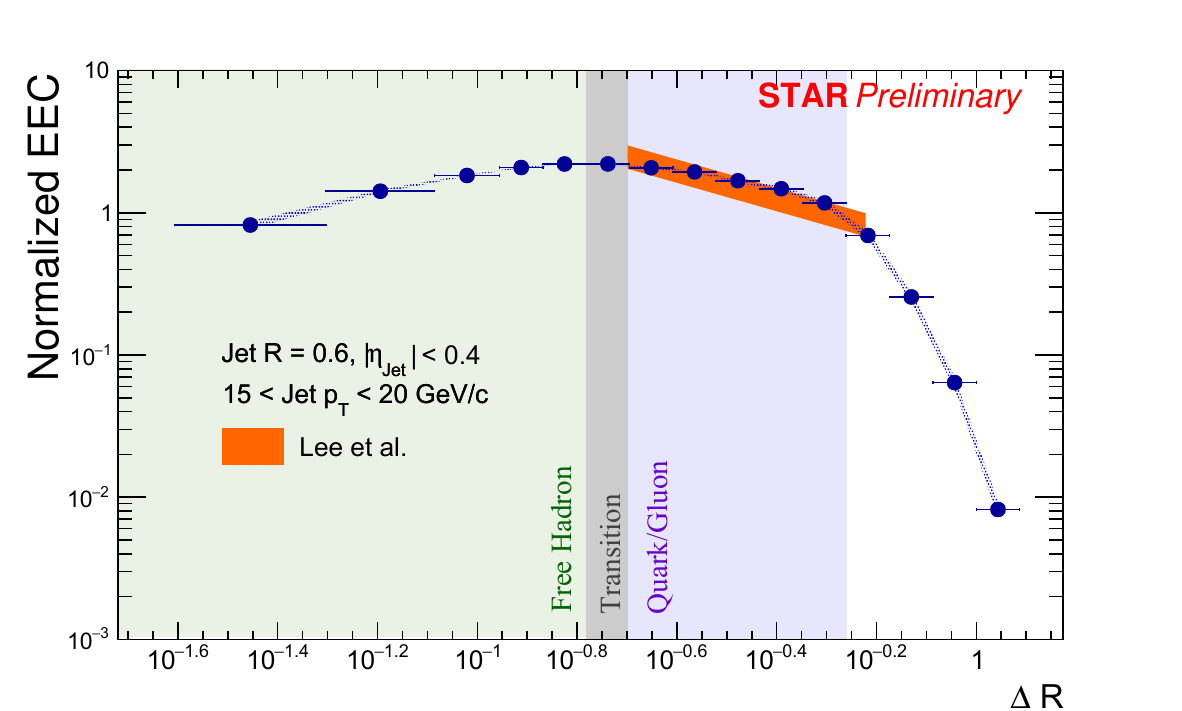}
\includegraphics[width=0.49\linewidth]{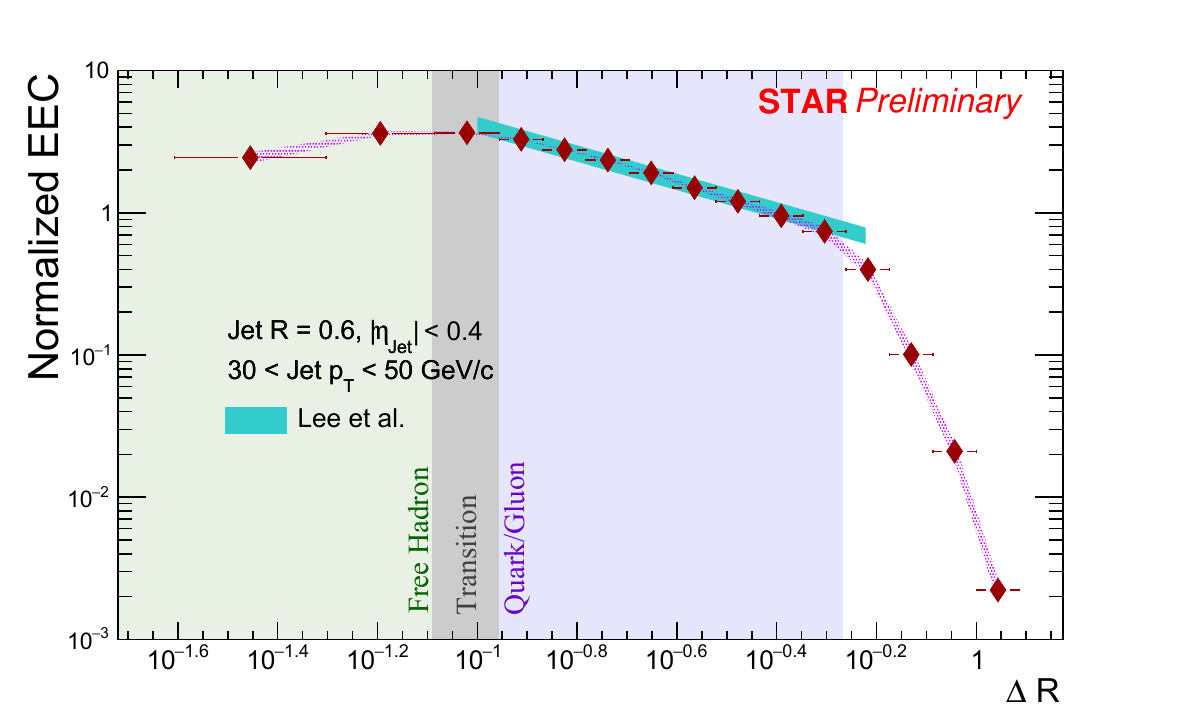}
\caption{Corrected distributions of the normalized EEC plotted differentially in $\Delta R$ for R = 0.4 (upper) and R = 0.6 (lower), for jet transverse momentum selections 15 < $p_{\rm T}$ < 20 GeV/c (left) and 30 < $p_{\rm T}$ < 50 GeV/c (right).  The free-hadron regime, transition region, and quark-and-gluon regime are highlighted in green, gray and purple respectively.  NLL-pQCD calculations are presented for ${\rm{3 GeV}}/{p_{\rm T,jet}}$ < $\Delta$ R < R.}
\label{Results}
\end{figure}

Fig. \ref{Results} shows the distribution of the EEC plotted on log-log scale for two selections of both jet radius and $p_{\rm T,jet}$.  The checkered bands indicate systematic uncertainties and the solid colored bands represent a Next-to-Leading-Log (NLL) perturbative QCD calculation \cite{Lee:2022ige}. At small angles, the results are consistent with a linear rise: reflecting the uniform distribution of energy associated with freely propagating hadrons \cite{Komiske:2022enw}.  The first bin that breaks this linear scaling behavior is labeled as the transition region, which is then followed by a decreasing cross section that behaves as approximately ${1}/{\Delta R}$.  The behavior in this region corresponds well with the theoretical calculation, indicating the ability of perturbative QCD to accurately describe the distribution of jet fragmentation at large $\Delta R$.  Finally, once the angular scale exceeds the radius selection of the jet cone, the distribution begins to fall off due to geometric limitations.  Additionally, comparing jet $p_{\rm T}$ selections, one can see that the transition region occurs at smaller $\Delta R$ for larger jet momentum, indicative of a later hadronization time.  Each particle pair is associated with a momentum transfer scale of $p_{\rm T,jet}  \Delta R$, which for the turn-over region is expected to occur at a constant value proportional to $\Lambda_{\rm{QCD}}$ \cite{Komiske:2022enw,Craft:2022kdo}.  This allows for the identification of a universal scale in angular distance that governs the breaking of perturbative behavior and the confinement of partons into hadrons that moves as a function of ${1}/{p_{\rm T,jet}}$.  This behavior is shown in the distributions shown in Fig. \ref{Results}, with the transition region occurring at a consistent value of $\Delta R_{\rm turn-over}  p_{\rm T,jet-low}$, where $\Delta R_{\rm turn-over}$ is computed using the lower and upper bounds of the transition region and $p_{\rm T,jet-low}$ is the lower bound of the jet transverse momentum selection.  This yields a value on the order of 2-3 GeV for three kinematic regions: 15 < $p_{\rm T,jet}$ < 20 GeV/c, 20 < $p_{\rm T,jet}$ < 30 GeV/c, and 30 < $p_{\rm T,jet}$ < 50 GeV/c.  Using this procedure, one can extract a similar value from the results in \cite{Komiske:2022enw}, implying consistency across a wide range of center-of-mass energy and jet momentum. Further studies will serve to constrain this value.

%\begin{center}
%\begin{tabular}{||c c c c||} 
% \hline
% $p_{\rm T,jet}$ [GeV/c] & $\Delta R_{turn-over} \textit{lower bound}$ & $\Delta R_{turn-over} \textit{ upper bound}$ & $\Delta R_{turn-over} * p_{T,jet-low}$\\ [0.5ex] 
% \hline\hline
% 15-20 & $e^{-1.8}$ & $e^{-1.6}$ & 2.5 - 3.0 \\ 
% \hline
% 20-30 & $e^{-2.2}$ & $e^{-2.0}$ & 2.2 - 2.7 \\
% \hline
% 30-50 & $e^{-2.5}$ & $e^{-2.2}$ & 2.5 - 3.3 \\
% \hline
%\end{tabular}
%\end{center}

\begin{figure}[ht]
\centering
\includegraphics[width=0.95\linewidth]{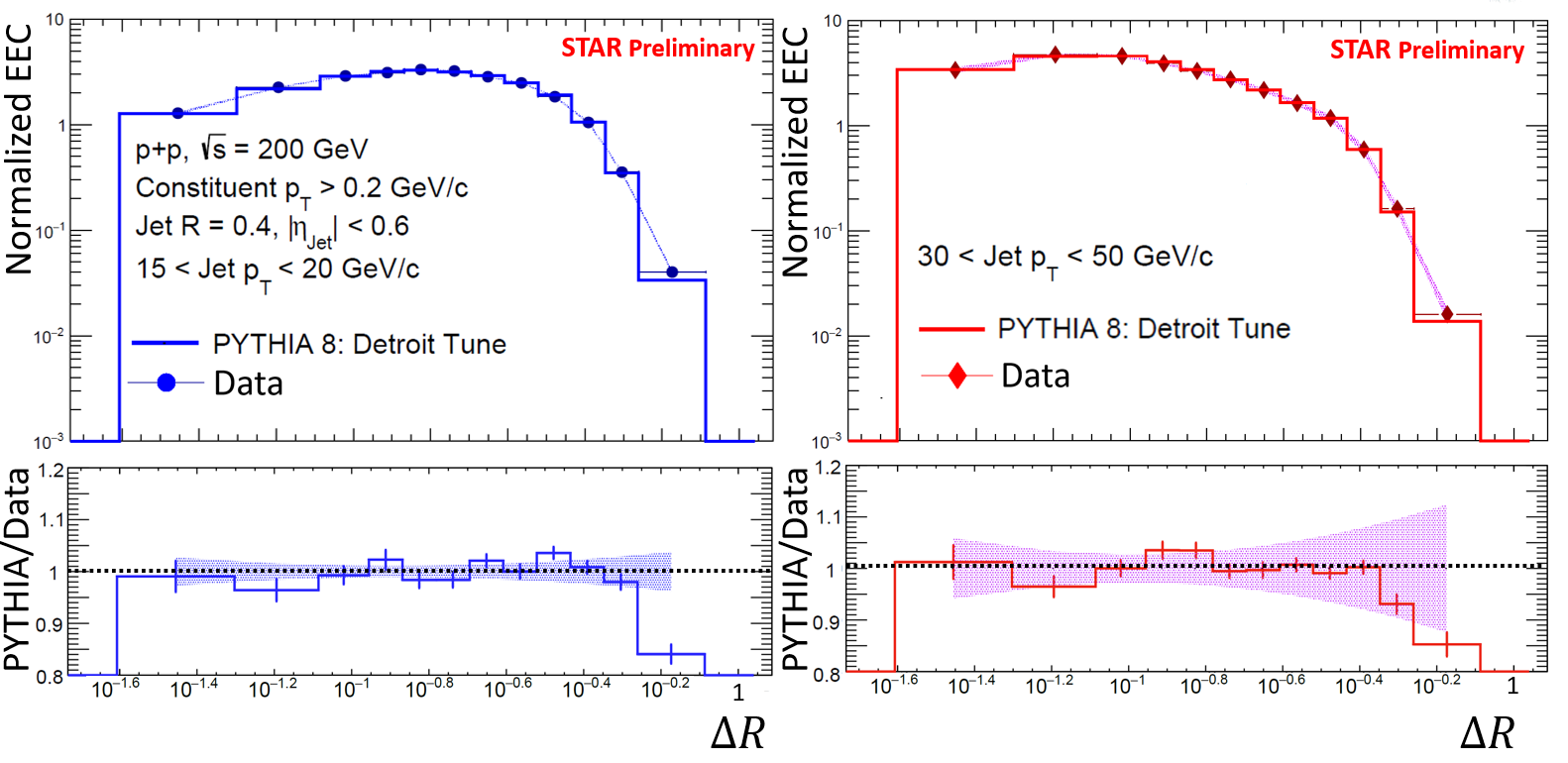}
\caption{Corrected distributions of the normalized EEC (top) plotted differentially in $\Delta R$ for R = 0.4, for jet transverse momentum selections 15 < $p_{\rm T}$ < 20 GeV/c (left) and 30 < $p_{\rm T}$ < 50 GeV/c (right).  Comparisons with PYTHIA-8 Detroit Tune are also presented.  The ratio of the PYTHIA distribution over the corrected data is also shown (bottom) alongside the magnitude of the systematic uncertainties for scale.}
\label{PYTHIA Comparison}
\end{figure}

Additionally, comparisons with the PYTHIA-8 Detroit Tune \cite{Aguilar:2021sfa} shown in Fig. \ref{PYTHIA Comparison} agree well within systematic uncertainties, with disagreement at large angles only occurring outside of the jet cone radius.  This shows that the energy flow of both freely diffusing hadrons and the perturbative shower of quarks and gluons are both captured accurately by PYTHIA simulations.  A well defined baseline of how the EEC behaves in vacuum will provide a useful reference for similar future studies done in heavy-ion collisions.

\clearpage
\section{Conclusions}

In these proceedings, the first corrected measurement of the two-point energy correlator in jets at RHIC is presented.  The EEC distribution has been shown to reveal the separation between three regimes: i) the free hadron phase at small opening angles, ii) perturbative behavior of quarks and gluons at large opening angles and iii) the transition region between them.  The behavior observed in the quark and gluon region agrees well with a NLL-pQCD calculation across jet momentum and jet radius selections.  Of particular interest is the transition region between the two distinct regimes of the correlator, which was observed to occur at an angle proportional to the inverse of jet transverse momentum.  This work will serve as the baseline for future measurements of the EEC in heavy-ion systems.  The time-proxy nature of this observable will allow for enhanced discrimination of when the jet is modified by the medium, by studying where the interaction is imprinted on the EEC distribution \cite{Andres:2023xwr}.

\end{document}